\documentclass[aps,prl,twocolumn,superscriptaddress]{revtex4-2}

\usepackage[colorlinks=true,citecolor=blue]{hyperref}
\usepackage{graphicx}
\usepackage{amsmath}
\usepackage{amssymb}
\usepackage{bbold}

\DeclareGraphicsExtensions{.png,.jpg,.eps}
\usepackage{xcolor}

\begin{document}

\author{Timo Hyart}
\affiliation{Department of Applied Physics, Aalto University, 00076 Aalto, Espoo, Finland}
\affiliation{International Research Centre MagTop, Institute of Physics, Polish Academy of Sciences, Al. Lotników 32/46, 02-668 Warsaw, Poland}

\author{J. L. Lado}
\affiliation{Department of Applied Physics, Aalto University, 00076 Aalto, Espoo, Finland}

\title{
Non-Hermitian many-body topological excitations in interacting quantum dots
}

\begin{abstract}
Quantum dots are one of the paradigmatic solid-state systems for quantum engineering, 
providing an outstanding tunability to explore fundamental quantum phenomena. Here we show that non-Hermitian many-body topological modes can be realized in a quantum dot chain by utilizing a gate-tunable modulation of dissipation, and they emerge purely because of the  non-Hermiticity. By exactly solving the non-Hermitian interacting description
both with exact diagonalization and tensor-networks, we demonstrate that these topological modes are robust even in the presence strong interactions, leading to a strongly correlated topological many-particle state.
Our results put forward quantum dot arrays as a platform for engineering non-Hermitian many-body
topological modes, and highlight the resilience of non-Hermitian topology
to electronic interactions.
\end{abstract}

\date{\today}

\maketitle

Non-Hermitian (NH) phenomena emerging in artificially designed systems
have motivated the rise of a new family of topological states \cite{Ashida2020,bergholtz2020exceptional}. 
On a mean-field level and in the case of specific types of Lindbladian dynamics, open quantum systems, experiencing gain and loss of energy due to coupling to the environment, can be described with effective NH Hamiltonians \cite{Ashida2020, El-Ganainy2018, bergholtz2020exceptional, Song2019, Lieu2020, Kawabata19,PhysRevB.103.L201114}.  Such approach has provided profound insights for the description of NH topological phases,  unconventional NH bulk-boundary correspondence and skin effect. Paradigmatic experiments probing the NH topological phases have mainly concentrated on photonic systems and electrical circuits
\cite{Ozawa, Zeuner2015, Poli2015, Zhan2017, Xiao2017, Weimann2017, Zhao2018, Bandres2018, Parto2018, Helbig2020}.
However, theoretical works have demonstrated that NH topological phases can be realized also in fermionic systems and superconductors, where the NH self-energy arises due to coupling to a reservoir 
\cite{Pikulin2012, Pikulin2013, Pikulin14, Bergholtz2019}. 

\begin{figure}[!t]
    \centering
    \includegraphics[width=0.96\columnwidth]{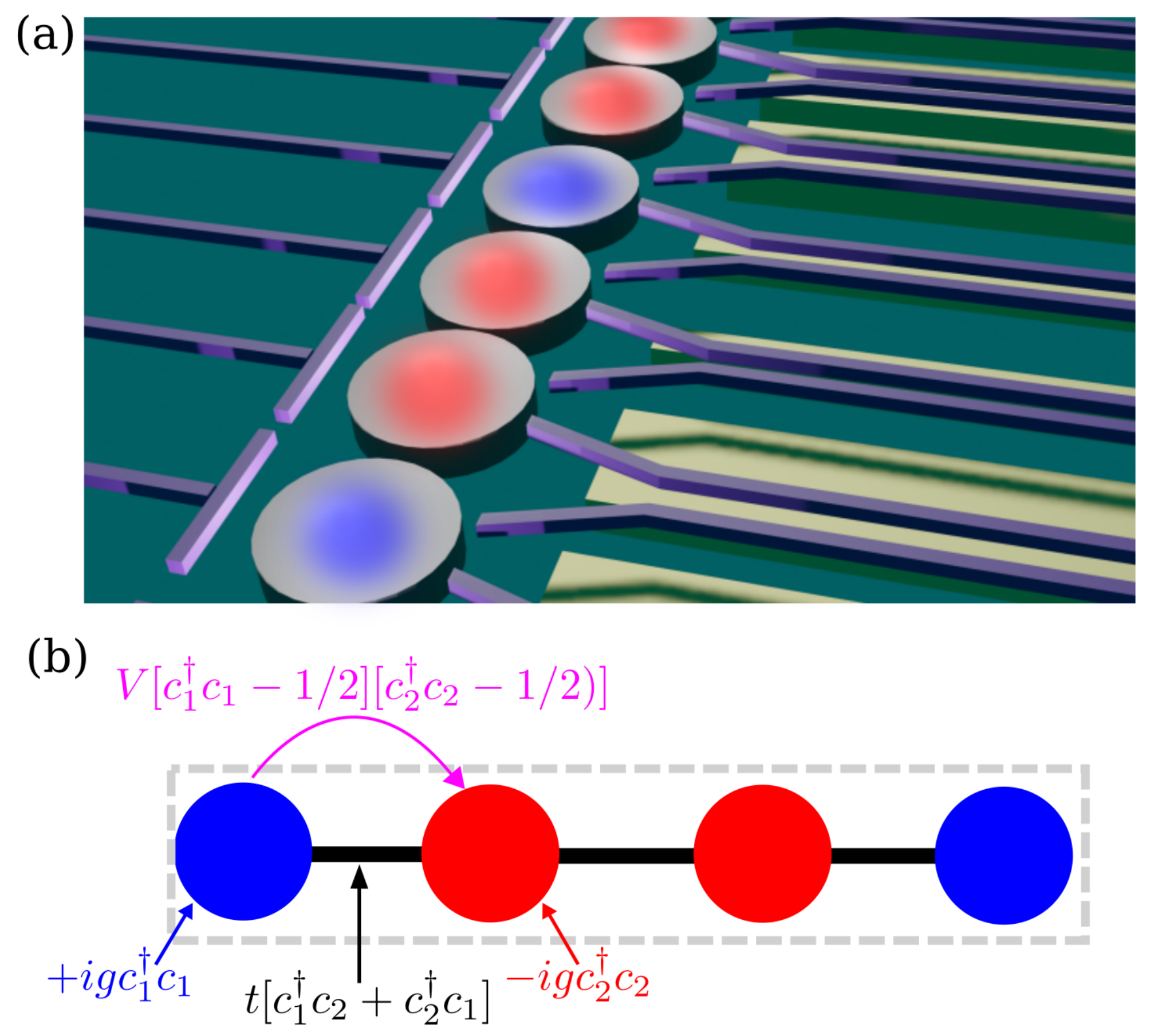}
	\caption{Schematic illustration of the tunable NH quantum dot chain. 
	The split-gate quantum point contacts control the NH self-energies $i g_i$ via the  tunnel couplings  between the quantum dots and the normal metal leads. The interaction strengths $V$ can be controlled either with the sizes of the quantum dots or by utilizing the screening of the normal metal leads. Additional gates can be used for tuning the energies of the quantum dot states and the tunneling $t$ between the quantum dots. 
}
    \label{fig:fig1}
\end{figure}

Here, we study interacting quantum dots as a simple platform to engineer a
correlated NH topological state. We show that 
an effective topological NH model can engineered
with the help of a modulation of the dissipation obtained by coupling the quantum dots to
normal metal leads via gate-tunable quantum point contacts (see Fig.~\ref{fig:fig1}). This type of tunable quantum dot arrays have been experimentally realized \cite{Hensgens2017, Mills2019, mortemousque2021enhanced} and previously considered for adaptive engineering of Hermitian topological phases \cite{Fulga_2013}.
Importantly, we demonstrate that the topological end modes of the NH state survive the presence of electronic
many-body interactions, by exactly solving the model with a NH tensor-network algorithm. 
Interestingly, the gate-tunability of the quantum dots allows to control effective parameters of the low-energy theory, including the interactions between the electrons and the dissipation. Thus, our results put forward quantum dot arrays as model systems for investigating the
correlated NH topology and demonstrating the resilience of the NH topology to many-body interactions.

The  NH model, tackled
in the quantum dot platform, is described by an effective
non-Hermitian Bloch Hamiltonian of the form
${\cal H}_0 = \sum_k c^\dagger_{k,\alpha} H_{k}^{\alpha,\beta} c_{k,\beta}$,
where
\begin{equation}
H_{k}=\begin{pmatrix}ig_{1} & t & 0 & te^{-ik}\\
t & ig_{2} & t & 0\\
0 & t & ig_{3} & t\\
te^{ik} & 0 & t & ig_{4}
\end{pmatrix},\label{eq:Ham}
\end{equation}
$t$ is the tunneling amplitude between the lattice sites and the non-trivial topology arises from the on-site non-Hermiticity $i g_i$ \cite{Takata2018,Wojtek19,Comaron20}.
While this model supports a family of topological
phases \cite{Takata2018,Wojtek19,Comaron20}, we focus on the case $g_1=g_4=0$ and 
$g_2=g_3=-2g$.  In this case, the topology is purely induced by dissipation and there is no net gain on any of the lattice sites. However, we point out that the model is topologically equivalent to a balanced gain-loss model $g_1=g_4=g$ and $g_2=g_3=-g$ [see Fig.~\ref{fig:fig1}(b)], because the latter is obtained from the previous one just by shifting the imaginary parts of the energies by a constant. We use these two models interchangeably but when calculating the observables, which are influenced by the lifetime of the excitations, we use the correct imaginary parts of the on-site energies \cite{supplementary}. 

\begin{figure}
    \centering
    \includegraphics[width=\columnwidth]{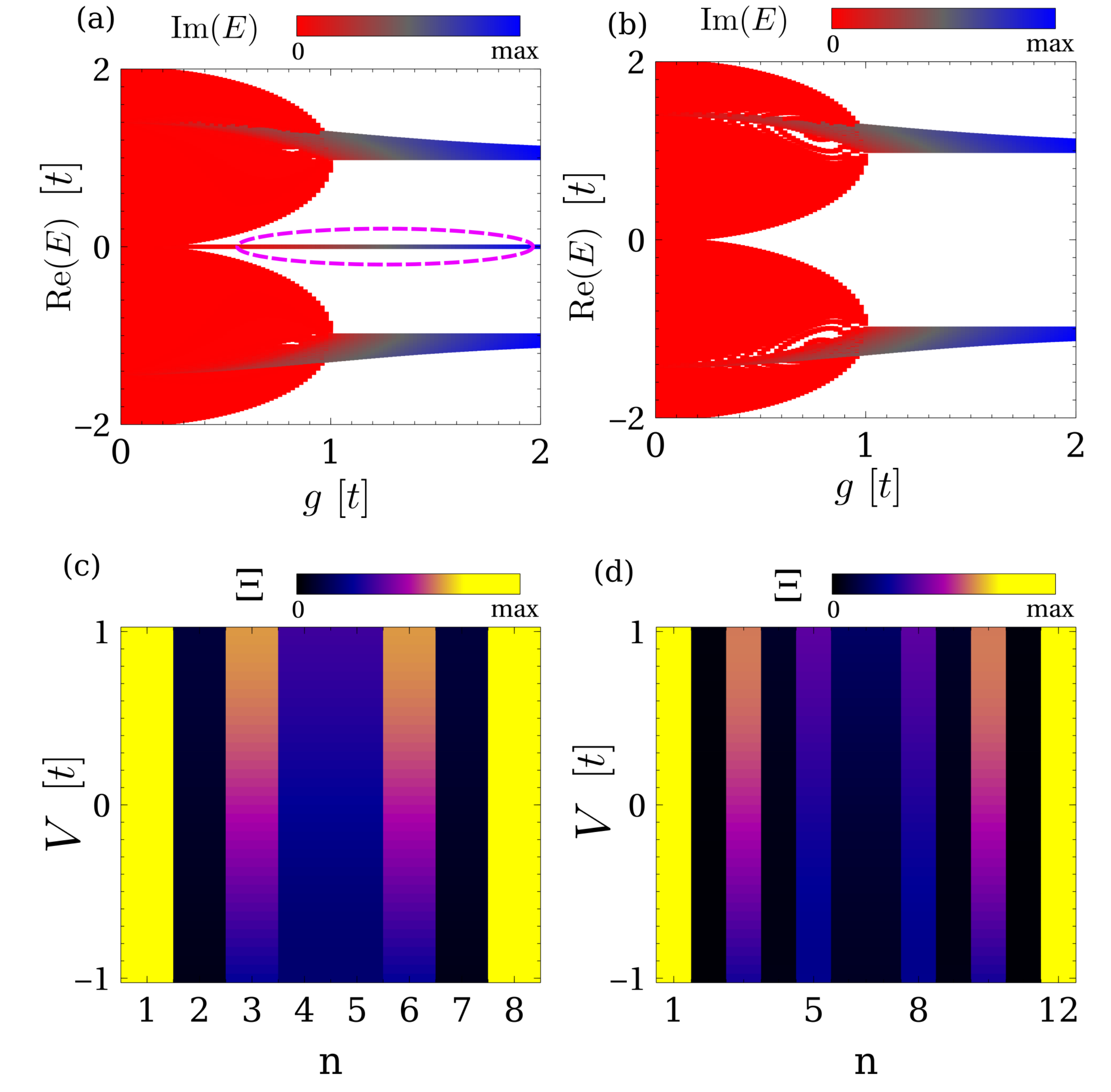}
	\caption{(a,b) Spectra for the non-interacting chain with (a) open and (b) periodic
	boundary conditions. The open chain supports topological end modes with zero real part of the energy. (c,d) Local correlator $\Xi (n)$, calculated using exact diagonalization of the interacting Hamiltonian, 
	as a function of the interaction strength $V$ demonstrating the
	resilience of zero modes to interactions. The chain lengths are $L=400$ in (a,b),
	 $L=8$ in (c)  and $L=12$ in (d).}
    \label{fig:fig2}
\end{figure}

The Hamiltonian $H_{k}$ satisfies a NH chiral symmetry ${\cal S} H_{k} {\cal S} = - H_{k}^\dag$ with ${\cal S}={\rm diag}(1,-1,1, -1)$, and therefore the topology of this 1D system is described by Chern number $C$ \cite{Wojtek19} in strong contrast to the Hermitian systems where the Chern number describes the topology of 2D systems \cite{Thouless82}. The Chern number is $C=-1$ for the parameters described above  giving rise to NH topological end modes with zero real part of the energy  \cite{Wojtek19, supplementary}. These end modes are shown in Fig.~\ref{fig:fig2}(a),(b) together with the bulk bands as a function of $g$. Although $C=-1$ for all values of $g$, increasing the dissipation leads to stronger localization of the zero modes to the end of the chain and, counterintuitively, increases the lifetime of these excitations \cite{supplementary}. For $g \sim t$ the localization length is few lattice sites enabling the study of these topological excitations in short chains. If not stated otherwise, we use $g=t$ in the numerical calculations presented below.

Quantum dot systems provide a natural way of realizing the model described above. Our starting point are the generic quantum dots, which do not possess any symmetries, so that spacing between the discrete energy levels is large and we can only take into account a single state in each quantum dot. Spin-degeneracy of the model
is broken by an applied magnetic field, and we focus in the regime in which tunneling amplitudes and interactions, and hence the topological gap, 
are smaller than the Zeeman energy scale. In typical semiconducting quantum dots a Zeeman splitting $\sim 1$ meV can be induced by applying a magnetic field $B=10$ T, so that we estimate that a  topological gap $\sim 0.1$ meV can be achieved in these systems by tuning the other parameters of the model. 
This is sufficient for the experimental study of the NH topology, but  we point out that much bigger topological gaps are possible in magnetic semiconductors and materials with large $g$-factor. The most important part of the model is, however, the site dependent non-Hermiticity, which can be achieved by utilizing a gate-tunable self-energy of the quantum dot state $i$ arising due to coupling to the normal metal lead  \cite{BeenakkerRMP, Datta-book, Flensberg-book}
\begin{equation}
\Sigma_i(E)=\lim_{\eta \to 0} \sum_{k} |t_{i,k}|^2 \frac{1}{E+i\eta-\epsilon_k}.
\end{equation}
Here $\epsilon_k$ is the eigenenergy of state $k$ in the lead and $t_{i,k}$ describes the tunneling amplitude from the quantum dot state $i$ to the state $k$ in the lead. The self-energy can be written as $\Sigma_i(E)= \tilde{E}_i - i g_i$, where $\tilde{E}_i$ renormalizes the energy of the quantum dot state $i$ due to the hybridization with the states in the lead and 
\begin{equation}
g_i(E)=\lim_{\eta \to 0} \sum_{k} |t_{i,k}|^2 \frac{\eta}{(E-\epsilon_k)^2+\eta^2}
\end{equation}
describes the finite lifetime of the electrons in the quantum dot. At low energies, $g_i(E)$ is independent of the energy and described by $g_i=\pi |t_{iN}|^2 \rho$, where $t_{iN}$ is the effective tunneling amplitude between the quantum dot and the lead and $\rho$ is the density of states in the lead \cite{Datta-book}. Thus, the dissipation $g_i$ in each quantum dot $i$ is sensitive to the voltage applied to a split-gate quantum point contact which determines the value of $t_{iN}$ (see Fig.~\ref{fig:fig1}).
Since the quantum dots are coupled to normal leads the energy and the number of particles are not conserved within the subsystem consisting of the quantum dot chain. Nevertheless, by applying gate voltages it is possible to maintain a constant average density of the electrons in the quantum dot system. Finally, we include the effects of electron-electron interactions. Within our low-energy model the dominating interaction term is
\begin{equation}
    \mathcal{H}_I =
    V \sum_n (c^\dagger_n c_n -1/2)(c^\dagger_{n+1} c_{n+1} -1/2), \label{interactions}
\end{equation}
where $V$ describes the strength of the electron-electron interaction between the  nearest-neighbor quantum dots. The Hamiltonian ${\cal H} ={\cal H}_0 + {\cal H}_I$, consisting of the non-interacting part ${\cal H}_0$ [Eq.~(\ref{eq:Ham})] and interactions ${\cal H}_I$ [Eq.~(\ref{interactions})] is the generic low-energy theory of the NH quantum dot chain. All parameters of the model are  tunable. The split-gate quantum point contacts control the NH self-energies $i g_i$, the interaction strengths $V$ can be controlled either with the sizes of the quantum dots or by utilizing the screening of the normal metal leads, and additional gates can be used for tuning the energies of the quantum dot states and the tunneling $t$ between the quantum dots.

We now move on to consider the effect of interactions in the previous model. In the case of interacting system we do not have direct access to the single-particle end state wave functions, but instead we have to express the quantities of interest using the many-particle wave functions. In the non-interacting case, there exists 4  many-particle states which all have the same real part of the energy up to the exponentially small corrections arising due to the finite size effects, and we can explicitly construct these states by considering the cases where each end mode is either occupied or unoccupied \cite{supplementary}. We find that this 4-fold degeneracy of the real parts of the energies survives in the presence of interactions.  Moreover, it resembles the topological degeneracies arising in many-particle Hermitian systems because it distinguishes the system with open boundary conditions from the system with periodic boundary conditions. In addition to the existence of the topological degeneracy we would also like to demonstrate that the many-particle excitations are localized at the end of the chain. For this purpose, we take into account also the exponentially small corrections in the real parts of the energies and denote the many-particle state with smallest real part of the energy as $|\Omega \rangle$ and the three other states with almost equal real parts of the energy as $|\Psi_i \rangle$ ($i=-1,0,1$), and we compute the local correlator defined as
\begin{equation}
    \Xi (n) = \sum_{i}
    |\langle \Psi_i | c_n | \Omega \rangle|^2.
\end{equation}
 \begin{figure}
    \centering
    \includegraphics[width=\columnwidth]{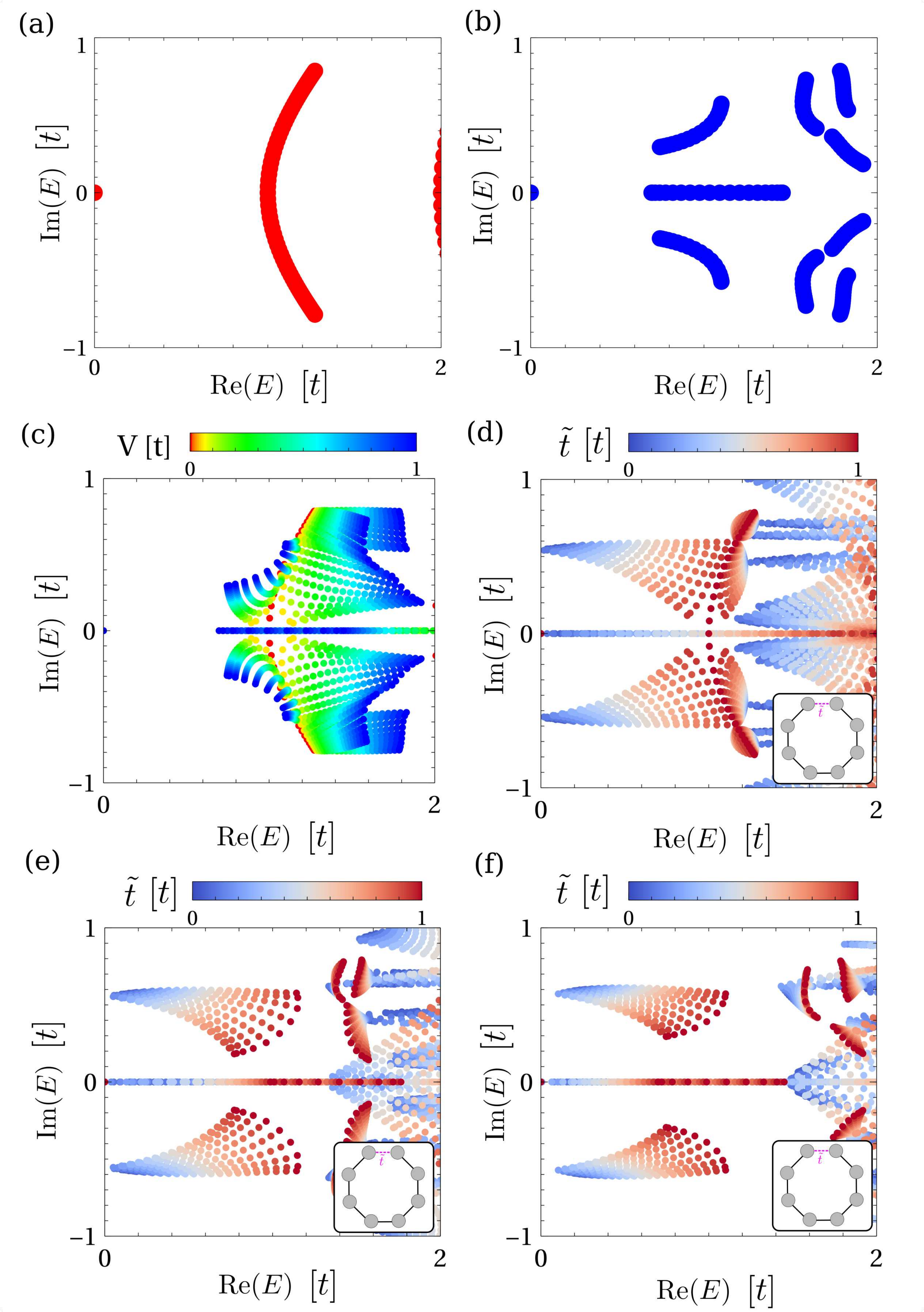}
	\caption{(a,b) Many-body energies relative to the lowest real part of the energy (averaged over twisted boundary conditions)
	for (a) $V=0$  and (b) $V=t$ with closed
	boundary conditions. (c) The evolution
	of the energies as $V$
	is ramped up. (d)-(f) The evolution of the energies
	as the boundary conditions are changed from
	closed ($\tilde t =t$) to open ($\tilde t =0$)
	for (d) $V=0$, (e) $V=0.5t$ and
	(f) $V=t$. We used $L=8$, but to simulate the energy bands we collected the energies corresponding to the different twisted boundary conditions.}
    \label{fig:fig3}
\end{figure}
In the non-interacting Hermitian systems, $\Xi (n)$ would yield the wave function of the topological end states. In particular, in a local tunneling 
experiment the zero-bias differential conductance would be determined by the local correlator 
$dI/dV (\omega=0,x) \sim \Xi (x)$. Due to these reasons, the local correlator $\Xi (n)$ is expected to yield information about the localization of the many-particle excitations. This indeed turns out to be the case, but it is worth pointing out that the analogies to the Hermitian systems are not perfect. First, the eigenstates of the non-interacting NH Hamiltonian are not orthogonal to each other. Due to this reason, the local correlator $\Xi (n)$ does not exactly describe the end state wave functions even in the non-interacting limit. Nevertheless, it is related to the end state wave functions and serves as a good indicator of the localization properties of the excitations \cite{supplementary}. Secondly, the excitations in NH systems have finite lifetimes, which show up as broadening of the peaks in the differential conductance. Therefore, the zero-bias differential conductance also includes a contribution from bulk states. Nevertheless, the bulk states are delocalized in the whole sample and therefore end excitations show up as a broadened zero-bias peak in the differential conductance in the vicinity of the end of the chain \cite{supplementary}. Therefore, $\Xi (n)$ as a function of $n$ serves as a good proxy for the local tunneling conductance profile.

Fig.~\ref{fig:fig2}(c),(d) show the local correlator $\Xi (n)$, obtained from exact diagonalization calculations, as a function of the interaction strength $V$ for short chains. The local correlator shows the emergence of topological end modes and demonstrates their resilience even in the presence of very strong interactions $V=t$, which leads to a full restructuring of the bulk bands as shown in Fig.~\ref{fig:fig3}(a),(b), where the many-body energies are measured relative to the lowest real part of the energy (averaged over twisted boundary conditions). To understand the robustness of the topological end modes, we notice that the bulk gap (line gap with respect to real part of the energies \cite{Kawabata19}) remains open upon increasing interactions from $V=0$ to $V=t$, demonstrating the existence an adiabatic connection between the
non-interacting and interacting limits [Fig.~\ref{fig:fig3}(c)]. We can further elaborate the correspondence between the end modes in the non-interacting and strongly interacting limits by varying one of the hopping amplitudes from closed ($\tilde t =t$) to open ($\tilde t =0$) boundary conditions [Fig.~\ref{fig:fig3}(d)-(f)]. The evolution of the spectrum as a function of $\tilde t$ is qualitatively similar for all values of $V$.
In the case of open boundary conditions, we observe the
emergence of the 4-fold many-body degeneracy in the real part of the energies both in the non-interacting and interacting limits, and therefore we interpret this degeneracy as the many-body signature of the topological phase. The imaginary parts of the  many-body energies are different due to the finite lifetime of the end modes.

\begin{figure}
    \centering
    \includegraphics[width=\columnwidth]{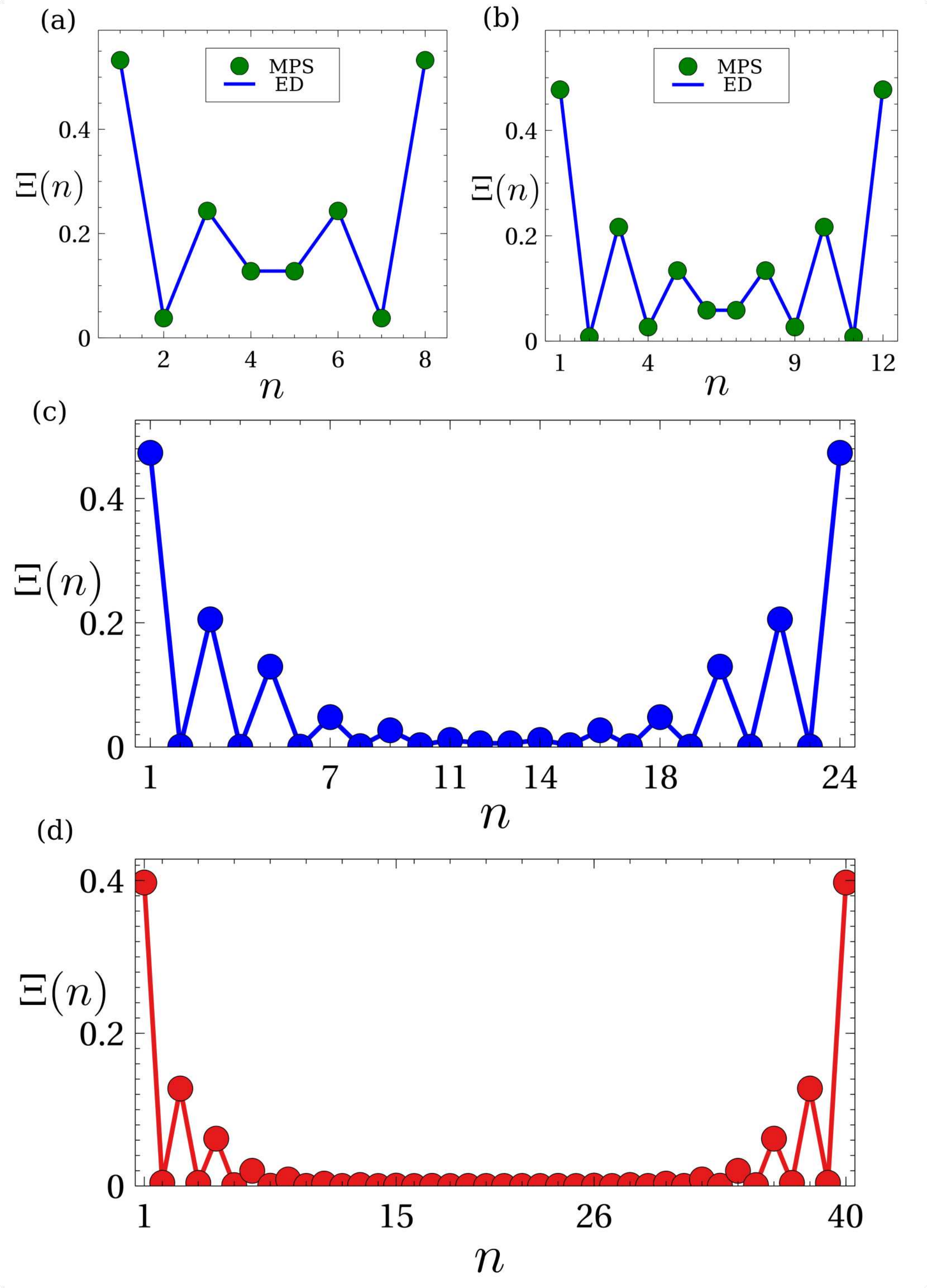}
	\caption{(a,b) Comparison between  $\Xi(n)$ obtained using exact diagonalization and the tensor-network formalism for short chains (a) $L=8$ and (b) $L=12$.
	(c,d) $\Xi(n)$  calculated using the tensor-network formalism 
	for long chains (c) $L=24$ and (d) $L=40$. The local correlator vanishes in the bulk implying that the low-energy excitations are localized at the end of the chain also in a strongly interacting system.
	We used $V=t$.}
    \label{fig:fig4}
\end{figure}

To get rid of the finite size effects, we have studied 
longer chains using a tensor-network formalism \cite{2020arXiv200714822F,ITensor,dmrgpy}. 
In our calculations, the NH
many-body eigenstates were represented as tensor networks, and we
implemented a NH Krylov subspace diagonalization \cite{arnoldi1951principle,Lehoucq1996,Tisseur2001,Stathopoulos1998,lehoucq1998arpack} targeting the many-body states with the lowest real parts of the energies.
To benchmark the algorithm, we show in Fig.~\ref{fig:fig4}(a,b) that for a short chain the local correlator obtained with exact diagonalization
and tensor-network formalism show a perfect agreement. In particular,
we observe the existence of end modes in the spatial profile of $\Xi(n)$. The important advantage of the tensor-network formalism is that we can solve the many-body eigenstates for much longer chains. The results for $L=20$ and $L=40$ are shown in Fig.~\ref{fig:fig4}(c,d).  They demonstrate that $\Xi(n)$ vanishes in the bulk, implying that the low-energy excitations are localized at the end of the chain also in a strongly interacting system with $V=t$. 

To summarize, we have proposed quantum dot chains as a viable experimental platform for studying interacting NH topological phases, and by performing calculations using exact diagonalization and
tensor-network algorithms, we have demonstrated that the topological end modes exists
in the presence of strong interactions and lead to a robust 4-fold topological degeneracy of the real parts of the many-particle energies. Our results put forward
a paradigmatic example of interacting topological matter, opening up interesting
 directions in the many-body NH physics. In particular,  we have demonstrated this phenomenology in a minimal interacting
NH model, whose
topological end modes arise from a non-trivial Chern number $|C|=1$. More generally, the Chern number is an integer topological invariant, and it is possible to construct models for free-fermion NH systems with arbitrarily large Chern numbers \cite{supplementary}. The topological classification of the interacting NH systems goes beyond the scope of this work, but we point out that the fate of the higher Chern number topological phases in interacting NH systems is a genuinely interesting problem for future research. On one hand, it is known that in Hermitian systems the interactions can lead to a collapse of the classification of the free-fermion systems \cite{Fidkowski1, Fidkowski2, Chiu2016}. On the other hand, in the case of $|C|=1$ we found an adiabatic continuity between the non-interacting and strongly interacting limits as the interactions were ramped up. In  interacting Hermitian systems, the many-body Chern number can be computed in various ways including the Green's function formalism \cite{Volovik2009, PhysRevB.83.085426, PhysRevX.2.031008, Wang2013,PhysRevB.100.125111,2021arXiv210302624L,PhysRevB.90.060502}, and assuming that the Green function at zero frequency satisfies the NH chiral symmetry this formalism generalizes to the family of interacting NH systems studied in our paper. Therefore, assuming that similar adiabatic continuity persists also in the case of higher Chern numbers, the topological end modes in the higher Chern number topological phases may also be resilient to strong interactions.

\begin{acknowledgments}
\textit{Acknowledgments:}
We acknowledge
the computational resources provided by
the Aalto Science-IT project.
J. L. L. acknowledges
financial support from the
Academy of Finland Projects No.
331342 and No. 336243. The research was also partially supported by the Foundation for Polish Science through the IRA Programme co-financed by EU within SG OP. 
\end{acknowledgments}

\bibliography{bibliography}


\onecolumngrid

\newpage

\setcounter{section}{0}
\renewcommand*{\thesection}{}
\section*{Supplementary material for "Non-Hermitian many-body topological excitations in interacting quantum dots"}

\section{Analytical results in the non-interacting limit}

In this section, we discuss the properties of the end states and bulk states of the non-interacting Hamiltonian 
\begin{equation}
H_{k}=\begin{pmatrix}ig & t & 0 & te^{-ik}\\
t & -ig & t & 0\\
0 & t & -ig & t\\
te^{ik} & 0 & t & ig
\end{pmatrix}.\label{eq:Ham-simple}
\end{equation}

\begin{figure}[h]
    \centering
    \includegraphics[width=0.9\columnwidth]{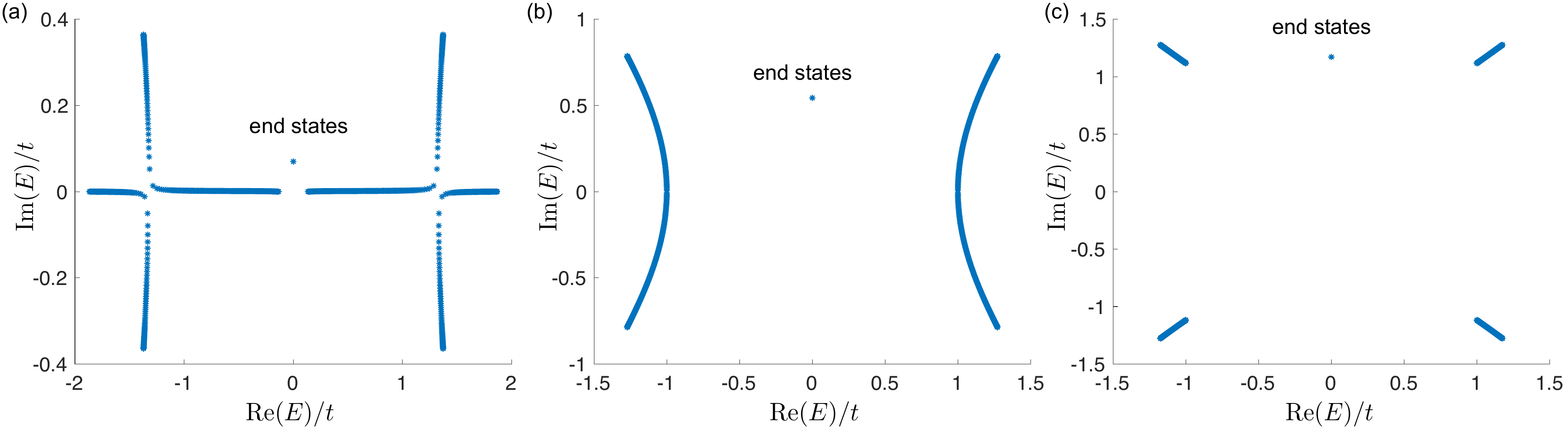}
	\caption{Spectrum for $L=800$ and (a) $g/t=0.5$, (b) $g/t=1$ and (c) $g/t=1.5$.}
    \label{fig:single-particle-spectrum}
\end{figure}

\begin{figure}[h]
    \centering
    \includegraphics[width=0.8\columnwidth]{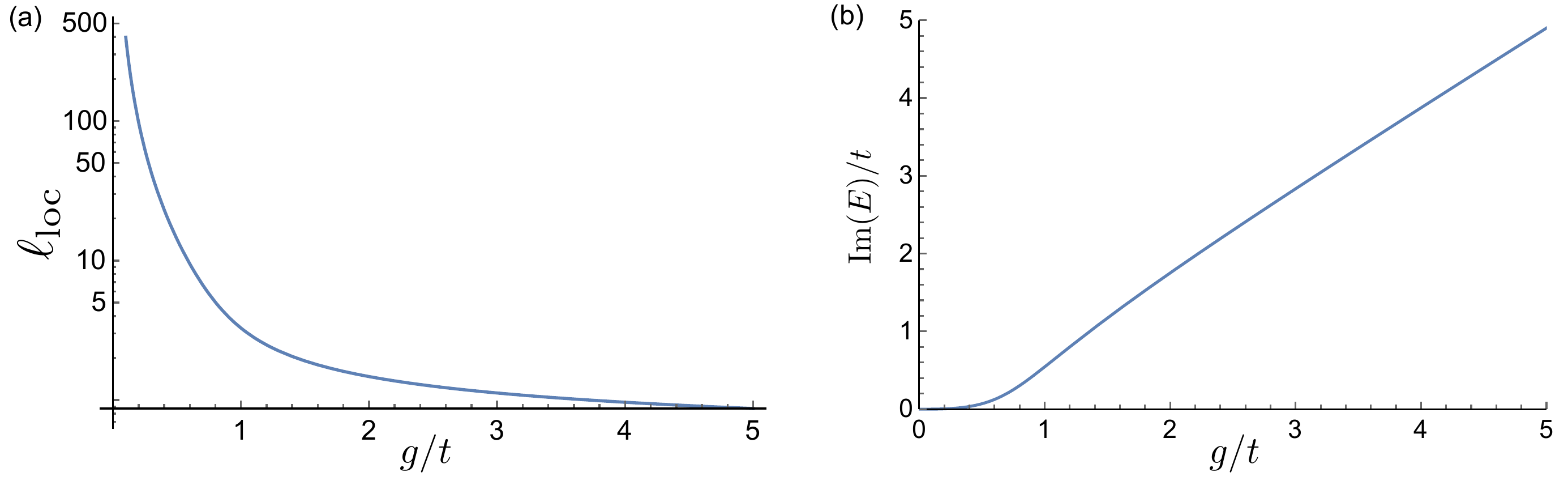}
	\caption{(a) Localization length of the end states $\ell_{\rm loc}$ as a function of $g/t$. For $g \sim t$ the localization length is already few lattice sites so that the topological end modes can be studied in small systems.  (b) The Imaginary part of the energy of the end states as a function of $g/t$.}
    \label{fig:endstateproperties}
\end{figure}

The single-particle bulk energies are given by
\begin{equation}
E=\pm t \sqrt{2-g^2/t^2 \pm 2 \sqrt{\cos^2(k/2)- g^2/t^2}}.
\end{equation}
There exists two qualitatively different bulk spectra for $g<t$ and $g>t$ as shown in Fig.~\ref{fig:single-particle-spectrum}, but in both cases the system supports topological end states. The wave functions and energies of the end states can be solved analytically. The full expressions are complicated but they can be used to calculate how the localization length $\ell_{\rm loc}$ and imaginary part of the energy of the end states depend on $g/t$. The results are shown in Fig.~\ref{fig:endstateproperties}. The real part of the energy of the end states is always zero.

\section{Topological invariant}

The topological invariant of the non-interacting model is the Chern number $C$ of an effective two-dimensional Hermitian Hamiltonian
\begin{equation}
H^{{\rm eff}}(k,\eta) \equiv \eta {\cal S} -i{\cal S} { H}_{k}=\begin{pmatrix}\eta+g & -it & 0 & -ite^{-ik}\\
it & -\eta+g & it & 0\\
0 & -it & \eta-g & -it\\
ite^{ik} & 0 & it & -\eta-g
\end{pmatrix},\label{eq:heffk}
\end{equation}
where ${\cal S}={\rm diag}(1,-1,1,-1)$ is the chiral symmetry operator. The Chern number can be calculated using the Kubo formula 
\begin{equation}
C=\frac{1}{2 \pi}\int_{-\infty}^{+\infty}d\eta\int_{0}^{2\pi}dk \ \Omega_{k,\eta},
\end{equation}
where the Berry curvature $\Omega_{k,\eta}$ is given by
\begin{equation}
\Omega_{k,\eta} = \sum_{{n\leq n_{F}\atop m>n_{F}}} {\rm Im}\frac{2\left\langle \psi_{k,\eta}^{n}\right|\!\partial_{k}H^{{\rm eff}}\!\left|\psi_{k,\eta}^{m}\right\rangle \!\left\langle \psi_{k,\eta}^{m}\right|\!\partial_{\eta}H^{{\rm eff}}\!\left|\psi_{k,\eta}^{n}\right\rangle }{\left(E_{k,\eta}^{(n)}-E_{k,\eta}^{(m)}\right)^{2}}.
\end{equation}
Here, $\left|\psi_{k,\eta}^{n}\right\rangle$ are the eigenstates and $E^{n}_{k, \eta}$ the eigenenergies of $H^{{\rm eff}}(k,\eta)$ (sorted in ascending order of eigenenergy), and $n_{F}$ is the number of occupied bands. By computing the Chern number we obtain $C=-1$.

\section{End state wave functions and local correlator in the absence of interactions}

In the non-interacting case, we can characterize the end states by solving their wave functions. On the other hand, in the interacting case we only have access to the many-particle wave functions, and in the main text we characterized the end excitations using a local correlator. In this section, we discuss the relationship of these concepts in the non-interacting limit. 

We can solve the single-particle wave functions $|\psi_i \rangle$ and eigenenergies $E_i$. We can order the wave functions based on increasing real part of $E_i$ and because our system supports a chiral symmetry  the energies satisfy ${\rm Re}(E_i)<0$ ($i=1,...,N$) and ${\rm Re}(E_i)>0$ ($i=N+1,...,2N$), where the system size is $L=2N$. Next we construct a many-particle wave function as
\begin{equation}
|\Omega \rangle = \frac{1}{\sqrt{{\cal N}}}\prod_{i=1}^N \gamma_i^\dag |0 \rangle, \ \gamma_i^\dag=\sum_n \psi_{i n} c_n^\dag.
\end{equation}
Here $\cal{N}$ is a normalization factor and $c_n^\dag$ is the fermionic creation operator at lattice site $n$. There exists a subtleity in the definition of this wave function because there can exist exceptional points where two eigenenergies become degenerate and the corresponding wave functions become parallel to each other. In the presence of these exceptional points the corresponding operator $\gamma_i$ should be included only once in the product. However, we point out that this removal is not necessary in our calculations, because even if the bulk Hamiltonian has exceptional points these degeneracies are typically removed due to the finite size effects. Nevertheless, the approximate degeneracies can show up as very small normalization factors, and in the limit of long chains the numerical calculations may become unstable.    

Importantly, because the end states have zero real part of the energy, there exists 3 other many-particle states which all have the same real part of the energy (we neglect the exponentially small corrections due to the finite size effects in the construction of these states)
\begin{equation}
|\Psi_{-1} \rangle = \frac{1}{\sqrt{{\cal N}_{-1}}}\prod_{i=1}^{N-1} \gamma_i^\dag |0 \rangle, \  |\Psi_{0} \rangle = \frac{1}{\sqrt{{\cal N}_{0}}}\prod_{i=1}^{N-1} \gamma_i^\dag \gamma_{N+1}^\dag |0 \rangle, \  |\Psi_{1} \rangle = \frac{1}{\sqrt{{\cal N}_{1}}}\prod_{i=1}^{N+1} \gamma_i^\dag |0 \rangle.
\end{equation}
As we demonstrate in the main text this 4-fold degeneracy of the real parts of the many-particle energies is a topological property which survives in the presence of interactions. Moreover, the presence of this additional degeneracy distinguishes the system with open boundary conditions from the system with periodic boundary conditions.

\begin{figure}[b]
    \centering
    \includegraphics[width=0.61\columnwidth]{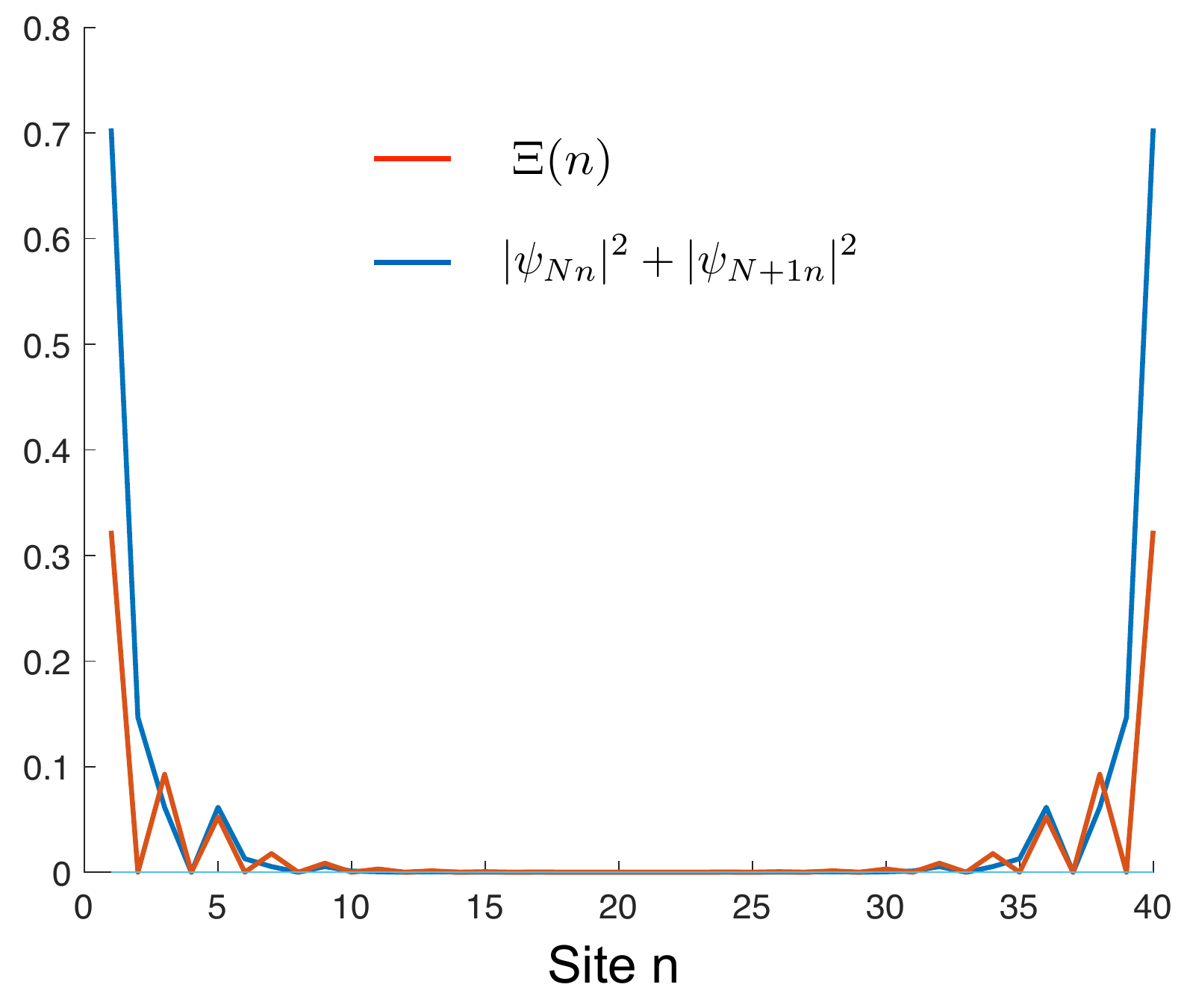}
	\caption{The end state wave functions $|\Psi_{Nn}|^2+|\Psi_{N+1 n}|^2$ and local correlator $\Xi (n)$ at the different lattice sites $n$. Here we have chosen $g=t$ and $L=40$.}
    \label{fig:WavefunctionCorrelator}
\end{figure}

In the absence of interactions we can characterize the two topological end states using the single-particle wave functions $|\psi_N\rangle$ and $|\psi_{N+1}\rangle$. However, in the presence of the interactions we have to characterize the end excitations using the many-particle wave functions $|\Omega\rangle$ and $|\Psi_i\rangle$ ($i=-1,0,1$) corresponding to the same real part of the total energy. We proposed to use a local correlator 
\begin{equation}
    \Xi (n) = \sum_{i=-1,0,1}
    |\langle \Psi_i | c_n | \Omega \rangle|^2
\end{equation}
for this purpose. It is easy to see that if we would be considering a non-interacting Hermitian problem the local correlator would be $\Xi(n)=|\psi_{N, n}|^2$ so that it would carry the same information as the single-particle end state wave functions. However, this relationship becomes more complicated in the non-Hermitian realm. Namely, by straightforward calculation we obtain that in the non-interacting limit $\Xi(n)$ is determined by
\begin{equation}
\Xi(n)= \frac{1}{{\cal N}_{-1} {\cal N}} \bigg| \det \begin{pmatrix} 
\langle \phi_n | \psi_1 \rangle & \langle \phi_n | \psi_2 \rangle &...& \langle \phi_n | \psi_N \rangle \\
\langle \psi_1 | \psi_1 \rangle & \langle \psi_1 | \psi_2 \rangle &...& \langle \psi_1 | \psi_N \rangle \\
\vdots & \vdots & ... & \vdots \\
\langle \psi_{N-1} | \psi_1 \rangle & \langle \psi_{N-1} | \psi_2 \rangle &...& \langle \psi_{N-1} | \psi_N \rangle
\end{pmatrix} \bigg|^2,
\end{equation}
where the inner products of the single-particle wave functions are defined as usual 
\begin{equation}
\langle \psi_i | \psi_j \rangle = \sum_n \psi_{in}^* \psi_{jm}    
\end{equation}
and the single-particle state $|\phi_n\rangle$ origanates from the operator $c_n$ and is therefore defined as $\phi_{nm}=\delta_{nm}$. The correlator and the end states wave functions are shown in Fig.~\ref{fig:WavefunctionCorrelator}. Both of them capture the presence of exponentially localized topological end modes, but the details of the spatial profile are slightly different. In particular the correlator vanishes at every second lattice site whereas the wavefunction is zero only at every fourth lattice site. 

\section{Tunneling conductance}

\begin{figure}[h]
    \centering
    \includegraphics[width=0.9\columnwidth]{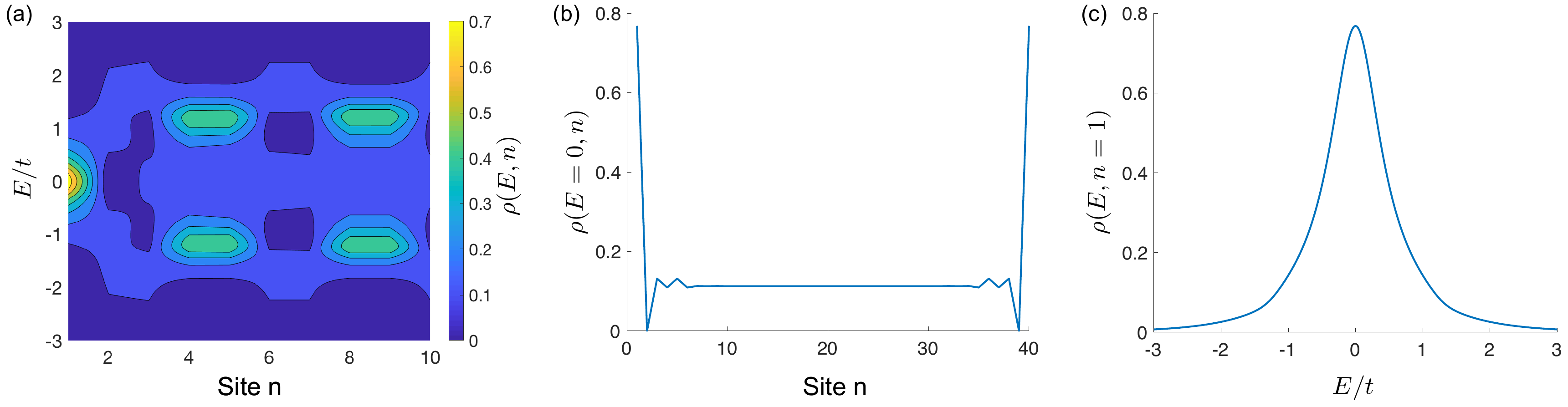}
	\caption{(a) The local density of states $\rho(E,n)$ close to the end of the chain as a function of the lattice site $n$ and energy $E$. (b) Linecut of $\rho(E,n)$ as a function of $n$ at zero energy $E=0$. (c) Linecut of $\rho(E,n)$ as a function of $E$ at the end of the chain $n=1$.  Here we have chosen $g=t$ and $L=40$.}
    \label{fig:densityofstates}
\end{figure}

Although in most calculations we can shift the Hamiltonian with a constant $i g \mathbb{1}$, so that we have a balanced gain-loss model, we have to invert the shift $\tilde{H}= H-i g \mathbb{1}$  when considering certain physical properties such as the tunneling conductance. This way we ensure that all the states have positive lifetimes. In the non-interacting limit the local density of states $\rho(E,n)$ at energy $E$ and lattice site $n$ can be calculated as 
\begin{equation}
\rho(E,n)=-\frac{1}{\pi} {\rm Im} \big[ \langle n| \frac{1}{E-\tilde{H}}| n \rangle \big].
\end{equation}
The topological end modes show up as a peak in the $\rho(E,n)$ at the end of the chain at energy $E=0$, where the width of the peak is determined by the lifetime of the end modes (see Fig.~\ref{fig:densityofstates}). The tunneling conductance $dI/dV$ at the lattice site $n$ and voltage $V$ is proportional to the $\rho(eV,n)$. 

\section{Models for topological phases with arbitrary large Chern number}

Topological phases with larger Chern numbers can be constructed by stacking the chains so that the Hamiltonian is
\begin{equation}
\bar{H}_{k}=\begin{pmatrix}
{ H}_{k}            & { H}_{T, k}       & 0                 & \dots                 & 0 \\
{ H}_{T, k}^\dag    & { H}_{k}          & { H}_{T, k}   & \ddots                &\vdots\\
0                       & { H}_{T, k}^\dag  & { H}_{k}      & \ddots                & 0 \\
\vdots                  & \ddots                & \ddots            & \ddots                & { H}_{T, k} \\ 
0                       & \dots                 & 0                 & { H}_{T, k}^\dag  & { H}_{k}
\end{pmatrix},\label{eq:Ham-stack}
\end{equation}
where ${ H}_{k}$ is given by Eq.~(\ref{eq:Ham-simple}) and 
\begin{equation}
{ H}_{T, k}=\begin{pmatrix}
0               & t_\perp       & 0         & 0\\
0               & 0             &t_\perp    & 0 \\
0               & 0             & 0         & t_\perp\\
t_\perp e^{ik}  & 0             & 0         & 0
\end{pmatrix}.\label{eq:Ham-tun}
\end{equation}
This Hamiltonian satisfies a chiral symmetry $\bar{{\cal S}}={\rm diag}({\cal S}, {\cal S}, {\cal S},...)$
and therefore for sufficiently small $t_\perp$ the Chern number is $C=-N_\perp$, where $N_\perp$ is the number of stacked chains. Numerical calculations demonstrate that the Hamiltonian (\ref{eq:Ham-stack}) indeed supports $N_\perp$ topological end modes at each end of the chain. The real parts of the energies of the end states are always zero. The coupling ${ H}_{T, k}$ between the chains can be easily realized by shifting each chain by one lattice site relative to the previous chain in the stack.

We point out that if one would not shift the chains relative to each other the tunneling Hamiltonian would be 
\begin{equation}
{ H}_{T, k}=\begin{pmatrix}
t_\perp         & 0              & 0         & 0\\
0               & t_\perp       & 0         & 0 \\
0               & 0             & t_\perp   & 0\\
0               & 0             & 0         & t_\perp
\end{pmatrix}.\label{eq:Ham-tun-simple}
\end{equation}
As a result the chiral symmetry takes a form $\bar{{\cal S}}={\rm diag}({\cal S}, -{\cal S}, {\cal S}, -{\cal S},...)$
so that for small values of $t_\perp$ the Chern number oscillates between $-1$ and $0$ as a function of $N_\perp$.

\end{document}